\documentclass[%
 reprint,
 amsmath,amssymb,
 aps,
]{revtex4-2}

\usepackage{amsmath}
\usepackage{graphicx}% Include figure files
\usepackage{dcolumn}% Align table columns on decimal point
\usepackage{bm}% bold math
\usepackage[utf8]{inputenc}
\usepackage[T1]{fontenc}
\usepackage{mathptmx}
\usepackage[T1]{fontenc} % Use modern font encodings
\usepackage{amsmath,color}
\usepackage{amssymb}
\usepackage{amsfonts}
\usepackage{amsthm}
\usepackage{mathtools}
\usepackage[title]{appendix}
\usepackage{hyperref}
\usepackage{braket}
\usepackage{xcolor}
\usepackage{etoolbox}
\usepackage{soul}
\usepackage{mathrsfs}
\usepackage[mathcal]{eucal}
\usepackage{tabularx,booktabs,array}
\newcolumntype{Y}{>{\centering\arraybackslash}X}

\newcommand{\argmin}{\operatornamewithlimits{argmin}}

\newcommand{\argmax}{\operatornamewithlimits{argmax}}

\usepackage{algorithm}
\usepackage{algpseudocode}
\usepackage{dsfont}

\usepackage{bbold}

\begin{document}

\title{\textit{Ab Initio} Free Energy Surfaces for Coupled Ion-Electron Transfer}
% Force line breaks with \\

\author{Ethan Abraham}
\email{ethana@mit.edu}
\affiliation{Department of Chemistry, Massachusetts Institute of Technology, Cambridge, Massachusetts 02139, USA}
\author{Martin Z. Bazant}
\affiliation{Department of Chemical Engineering, Massachusetts Institute of Technology, Cambridge, Massachusetts 02139, USA}
\affiliation{Department of Mathematics, Massachusetts Institute of Technology, Cambridge, Massachusetts 02139, USA}
\author{Troy Van Voorhis}
\affiliation{\mbox{Department of Chemistry, Massachusetts Institute of Technology, Cambridge, Massachusetts 02139, USA}}
\date{First posted 21 October 2025; revised 25 November 2026}% It is always \today, today,
\begin{abstract}
\vspace{1 cm} Although coupled ion-electron transfer (CIET) has emerged as a powerful framework to rationalize the kinetics of Faradaic reactions, its mechanism has lacked a fully microscopic, first-principles description. Here we show that the coupling of electron-transfer to ion-transfer can be understood as an extension of Marcus theory in which the ensemble of diabatic nuclear configurations is conditioned on a classical collective variable describing the interfacial anisotropy. This formalism enables direct construction of the CIET free-energy surface from constrained \textit{ab initio} trajectories, providing a first-principles route to electrochemical current-overpotential relations. We demonstrate this method for CO$_2$ redox on a gold electrode and find that the resulting two-dimensional saddle-point barriers differ substantially from one-dimensional treatments that consider only electron- or ion-transfer coordinates individually.
\end{abstract}
 \maketitle

\makeatletter
\begingroup
\renewcommand\@makefntext[1]{\noindent#1}% no footnote mark
\def\@thefnmark{}%
\interfootnotelinepenalty=10000 % keep the note in one block
\@footnotetext{This version incorporates the corrections published in Erratum, Phys. Rev. Lett. 137, 229901 (2026), doi:10.1103/h8h2-hsbs. Tables I and II and Figs. 1(b), 2(a), and 2(b) differ from the originally published Letter (see Erratum for details); the formalism, equations, and stated conclusions are unchanged. The processed grids are now available at \url{https://github.com/eabes23/ab_initio_ciet/tree/15d5a12}.}
\endgroup
\makeatother

Marcus free energy curves relate macroscopically observed electron transfer (ET) reaction rates to a microscopic picture of solvent fluctuations \cite{Marcus_1956,Marcus_1964,Hush_adiabatic}. The iconic parabolic free energy curves that describe the ET landscape arise from the fact that the energy gap between the reactant and product diabatic electronic states is mediated by the instantaneous nuclear configurations of the solvation environment, and that the probability distribution of such energy gaps is Gaussian to a good approximation; taking the logarithm of this distribution yields the parabolas \cite{Tachiya_1989, Warshel_Intro, Blumberger_review}. Crucially, this framework assumes that nuclear configurations are sampled from a single, global ensemble---a simplification that may break down in anisotropic environments such as electrochemical interfaces or biomolecular structures \cite{Bazant_CIET,early_CIET,Nocera_PCET,Schiffer_PCET}.

In this Letter, we show that in the presence of such anisotropy, ET can become coupled to a classical collective variable (CV) that characterizes the broken symmetry. This is suggested by the fact that the nuclear ensembles, when conditioned on different values of the CV, are generally distinct. As a result, the Marcus curves associated with each value may differ in curvature, vertical shift, and electronic coupling. The reaction landscape is thus more appropriately viewed as a 2D manifold of Marcus curves parametrized by the classical CV.

Extensions of Marcus theory to cases in which ET is coupled to another degree of freedom have already found applications. Proton-coupled electron transfer (PCET) generalizes Marcus theory by extending the reaction coordinate to include a proton degree of freedom \cite{Nocera_PCET,Schiffer_PCET}. However, the proton is typically treated quantum mechanically and associated with a reaction coordinate involving vibronic excitations rather than a classical CV. More recently, coupled ion-electron transfer (CIET) has emerged as a framework to rationalize the rates of redox reactions in commercial lithium-ion battery electrodes \cite{Stenlid_2024,LIB_Science}. In this theory, the Marcus ET coordinate is coupled to a classical ion transfer (IT) coordinate related to the distance from the electrode-electrolyte interface, and the resulting rate expressions fit the experimentally observed current-overpotential curves significantly better than rate expressions that do not take into account ion-electron coupling \cite{LFP_Images,LIB_Science,Bazant_CIET}. Unlike methods which use a single ground-state pathway determination to compute the barrier, CIET requires integration over such paths for a band of electrode states in order to directly derive the current-overpotential relationship \cite{Abraham2025_lambda_eff,Bazant_CIET,Yoon_2026}. 

We note that, although Marcus theory provides the standard microscopic picture of ET reactions, the Butler–Volmer (BV) equation has empirically described Tafel-like current–voltage relations for reaction-limited electrochemical currents for nearly a century \cite{Gurney_1931,Butler_1936}. In its usual form, however, BV corresponds to effectively linear diabatic free-energy curves, whereas Marcus theory is based on quadratic free-energy surfaces \cite{Marcus_1956,Marcus_1964}. CIET resolves this apparent discrepancy by postulating a two-dimensional free-energy surface which, in the simplest approximation, is quadratic along the Marcus ET coordinate and linear along the ion-transfer collective variable \cite{Bazant_CIET}. As a result, the CIET rate expression reduces to Marcus–Hush–Chidsey (MHC) or Butler–Volmer-like kinetics in the appropriate limits, depending on which direction in the landscape controls the barrier \cite{Bazant_CIET,Abraham2025_lambda_eff,MHC_Simple,Chidsey}. This provides a physical basis for why different kinetic expressions succeed in different electrochemical regimes and suggests that conditioning ET on a symmetry-breaking classical coordinate offers a general extension of Marcus theory for anisotropic condensed-phase environments \cite{Nitzan_2006,Monte_carlo_charge_transfer}.

While the success of CIET in phenomenologically describing Faradaic kinetics has been promising \cite{CIET_LCO,LFP_Images,Bazant_CIET,Yoon_2026}, no microscopic formulation of CIET amenable to first principles calculation has been reported. Although \textit{ab initio} calculation of CIET parameters has recently been explored in Ref. \cite{Stenlid_2024}, the framework accounted neither for finite-temperature fluctuations \cite{Tachiya_1989,VanVoorhis_2010,Blumberger2006} nor the effect of the electronic coupling on the free energy surface \cite{Abraham2025_lambda_eff} (see Supplementary Material \cite{SM}). Here we present a physically consistent formalism to compute finite-temperature free-energy surfaces for ET coupled to a classical CV, expressed entirely in terms of quantities derived from constrained \textit{ab initio} trajectories. By computing the transition-state (TS) of the CIET surface as a function of the thermodynamic driving force, our method enables the calculation of current--overpotential curves from first principles and advances the prospect of \textit{ab initio} electrochemical kinetics. Importantly, we note that practical analyses using these multidimensional surfaces are expected to become more feasible with concurrent improvements in sampling methods and machine-learned potentials \cite{jarzynski1997,bussi2006equilibrium,barducci2008well,Musaelian2023Allegro,batzner2022e3,unke2021machine}.

{\it Constructing the CIET Surface in the Diabatic Representation.} Within the Born–Oppenheimer approximation, an \textit{ab initio} trajectory 
constrained to a diabatic electronic state can be modeled using constrained 
density functional theory (cDFT) \cite{Troy_cDFT_proof,VanVoorhis_2010,VanVoorhis_2012,Troy_original_cdft}, as implemented in 
various software packages \cite{cdft_cp2k,Holmberg_cdft_Implement,cdft_projector}. 
Consider a simulation of $N$ atoms with the electronic state fixed to 
$\ket{\alpha} = \ket{\alpha(\mathbf R)}$, where 
$\mathbf R \in \mathbb R^{3N}$ denotes the nuclear configuration. 
Let $\mathbf R_\alpha(t)$ denote a nuclear trajectory under this constraint. 
At steady state, the probability of finding the system in configuration 
$\mathbf R$ is given by the time-independent distribution $p_\alpha(\mathbf R)$. 
Letting $\alpha = O,R$ denote the initial and final (e.g., oxidized and reduced) diabatic states of the electron 
transfer, the ET coordinate of $\mathbf R$ is defined by the energy gap \cite{Tachiya_1989,Warshel_1982,Warshel_2001}
\begin{equation}\label{gap}
q(\textbf{R}) = \bra{O(\textbf{R})}\hat{H}\ket{O(\textbf{R})}-\bra{R(\textbf{R})}\hat{H}\ket{R(\textbf{R})},
\end{equation} where $\hat{H}$ is the Hamiltonian. Now suppose $\xi=\xi(\textbf{R})$ is a CV that characterizes the anisotropy. Then the histogram $\mathscr{N}_\alpha(q,\xi)$ is obtained from the trajectory by integrating over the appropriate region of the phase space \begin{equation}\label{prob}
\mathscr{N}_\alpha(q,\xi)=\int dt\int d^{3N}\textbf{R}\delta(q(\textbf{R}_\alpha(t))-q)\delta(\xi(\textbf{R}_\alpha(t))-\xi).
\end{equation}
The probability distribution is then given by $p_\alpha(q, \xi) = \mathscr{N}_\alpha(q, \xi)/\int dqd\xi\mathscr{N}_\alpha(q, \xi)$, and the 2D \textit{excess} free energy \footnote{Formally, the free energy used in this work is the excess free energy $G^\text{ex}$, which retains only the terms that affect the potential experienced by an individual particle, subtracting off the ``counting'' effect related to concentration \cite{Bazant_CIET}. However, we denote this quantity by the standard free energy symbol $G$ for notational simplicity.} surface is given by
\begin{equation}\label{gibbs}
G_\alpha^\text{ex}(q,\xi)=-k_BT\text{ln}p_\alpha(q,\xi)+C_\alpha,
\end{equation}
where $C_\alpha$ is a constant. It is convenient when applicable to define the equilibrium coordinates for each state \begin{equation}\label{bounds}(q_\alpha,\xi_\alpha)=\argmax_{q,\xi}p_\alpha(q,\xi).\end{equation} Then, noting that the constants $C_O$ and $C_R$ must be related such that the reaction free energy between these equilibrium states is the thermodynamic driving force $\Delta G_{OR}^\text{ex}$, a natural choice is \begin{subequations}\label{const}\begin{align}C_O &= k_BT \ln p_O(q_O,\xi_O), \label{eq:Ci} \\
C_R &= \Delta G_{OR}^\text{ex} + k_BT \ln p_R(q_R,\xi_R) \label{eq:Cf}
\end{align}\end{subequations} so that the equilibrium reactant state has $G_O^\text{ex}(q_O,\xi_O)=0$.

A practical challenge is that in the presence of high reaction barriers, the sampled histogram $p_\alpha(q,\xi)$ is often unpopulated for many regions in the 2D reaction coordinate space including, crucially, near the TS. However, this challenge can easily be overcome by employing methods common in Marcus theory (for $q$) and classical enhanced sampling (for $\xi$). Sampling over the CV can be achieved by standard techniques such as metadynamics or harmonic restraints \cite{constrained_pmf,jarzynski1997,Bussi2006}. If $U^{\text{bias}}_\alpha(\xi)$ is the resulting bias potential, the true distribution $p_\alpha(q,\xi)$ can be obtained from the observed distribution via \begin{equation}\label{pmf}p_\alpha(q,\xi)=p^{\text{obs}}_\alpha(q,\xi)\text{exp}[+U^{\text{bias}}_\alpha(\xi)/k_BT].\end{equation} Once $\xi$ is sampled, the cross-sectional Marcus free energy curves are obtained by invoking the standard assumption that they are parabolic, whereby
\begin{subequations}\label{2dmarcus}\begin{equation}p_{\alpha}(q|\xi)=\frac{1}{\sqrt{4\pi k_BT\lambda_{\alpha}(\xi)}}\text{exp}\left(-\frac{(q-\text{E}_\alpha(q|\xi))^2}{4k_BT\lambda_{\alpha}(\xi)}\right),\label{distribution}\end{equation}\begin{equation}\lambda_{\alpha}(\xi)=\frac{\text{Var}_\alpha(q|\xi)}{2k_BT},\label{lambda} 
\end{equation}\end{subequations} or under the assumption that the ET is symmetric,
\begin{equation}\label{symmetric_lambda}\lambda_\alpha(\xi)=\lambda(\xi)=\frac{1}{2}(\text{E}_R(q|\xi)-\text{E}_O(q|\xi)).\end{equation}
Here $\text{E}_\alpha(q|\xi)$ and $\text{Var}_\alpha(q|\xi)$ denote, respectively, the mean and variance of the conditional distribution $p_{\alpha}(q|\xi),$ which can be approximated from $p^{\text{obs}}_{\alpha}(q|\xi)$ \footnote{Interestingly, if we assume the energy gap sampling (Eq. (\ref{gap})) is harmonic, equating the change in free energy between the minima of the Marcus parabola at a particular value of the CV implies that $U^{\text{bias}}_R(\xi)-U^{\text{bias}}_O(\xi)=
\frac{1}{2}(\text{E}_O(q|\xi)+\text{E}_R(q|\xi)-\lambda_R(\xi)+\lambda_O(\xi)).$ This equation relates the ET-coordinate statistics to the CV statistics, although in practice, it may not hold exactly due to asymmetry/anharmonicity in the ET, sampling insufficiency, or methodological uncertainty.}.

In the standard case where the sampling is performed such that $p^{\text{obs}}_{\alpha}(\xi)=\int dqp^{\text{obs}}_{\alpha}(q,\xi)$ is constant, using Eq. (\ref{const})-(\ref{2dmarcus}), we now obtain for our free energy surfaces
\begin{subequations}\label{result}\begin{equation} G_O^\text{ex}(q,\xi)=-[U^{\text{bias}}_O(\xi)-U^{\text{bias}}_O(\xi_O)]+\frac{(q-\text{E}_O(q|\xi))^2}{4\lambda_O(\xi)},\end{equation}\begin{equation}G_R^\text{ex}(q,\xi,\Delta G_{OR}^\text{ex})=-[U^{\text{bias}}_R(\xi)-U^{\text{bias}}_R(\xi_R)]+\frac{(q-\text{E}_R(q|\xi))^2}{4\lambda_R(\xi)}+\Delta G_{OR}^\text{ex}.\end{equation}\end{subequations} (We assume that both the ET and CV coordinates are overdamped, with the CV relaxing on a slower timescale than the ET coordinate, allowing $p_\alpha(q|\xi)$ to be obtained from enhanced sampling \cite{TS_review,Nitzan_2006}. These assumptions will be revisited in future work).

Finally, the free energy curves obtained via the above method can be used to compute the TS and activation barrier. The intersection of the curves is given by the set
\begin{equation}\label{inter}
\begin{aligned}
\mathcal{I}=\{(q,\xi)~:~G_R^\text{ex}(\xi,q)=G_O^\text{ex}(q,\xi)\},
\end{aligned}
\end{equation} and the TS and activation barrier in the nonadiabaitc limit are given by
\begin{subequations}\label{ts}\begin{align}
(\xi_\ddagger,q_\ddagger)&=\argmin_{(q,\xi)\in\mathcal{I}}~G_O^\text{ex}(q,\xi),\\ 
\Delta G_\ddagger^\text{ex}&=G_O^\text{ex}(q_\ddagger,\xi_\ddagger).
\end{align}\end{subequations}

For visualization and comparison of landscapes between systems, it is convenient to nondimensionalize both reaction coordinates. For the CV, the natural choice is \begin{equation}\label{CVnondim}
\tilde{\xi}=\frac{\xi-\xi_O}{\xi_R-\xi_O},
\end{equation} where $\xi_O$ and $\xi_R$ are determined from Eq. (\ref{bounds}) and the tilde denotes a nondimensionalized quantity. If we plot the CV and ET coordinates on perpendicular axes (see Supplementary Material \cite{SM} for analysis of the coordinate orthogonality), and enforce that the initial and final Marcus parabolas in each cross section lie at 0 and 1 respectively, we obtain for the ET coordinate \begin{equation}\label{ETnondim}
\tilde{q}(\tilde{\xi})=\frac{q-\text{E}_O(q|\tilde{\xi})}{\text{E}_R(q|\tilde{\xi})-\text{E}_O(q|\tilde{\xi})}.
\end{equation} Note that the nondimensionalization scheme stretches the ET coordinate differently at different CV values. However, it can be easily seen that this transformation maps only the ranges of the reaction coordinates, and so barrier heights in Eq. (\ref{ts}b) are unaffected by the non-dimensionalization.

Once obtained, the activation barrier can be used to derive ET rate expressions, providing an \textit{ab initio} route toward current-overpotential relations. As in standard Marcus theory, we note that the thermodynamic free energy $\Delta G_{OR}^\text{ex},$ which determines the shift of the two diabatic surfaces relative to each other, can depend on factors such as electrochemical overpotential, which only weakly affect the shape of the individual surfaces. Thus $\Delta G_\ddagger^\text{ex}=\Delta G_\ddagger^\text{ex}(\Delta G_{OR}^\text{ex})$ can be obtained from Eq. (\ref{ts}) numerically, and the nonadiabatic rate expression for a discrete transition can be determined from Fermi's golden rule
\begin{subequations}\label{golden}\begin{align}
k(\Delta G_{OR}^\text{ex}) &= \frac{2\pi}{\hbar} \, |H_{OR}|^2 \, \mathscr{P}(\Delta G_{OR}^\text{ex}),\\\mathscr{P}(\Delta G_{OR}^\text{ex})&=Q^{-1}\text{exp}[-\Delta G_\ddagger^\text{ex}(\Delta G_{OR}^\text{ex})/k_BT],
\end{align}\end{subequations} where $Q=\int d\Delta G_{OR}^\text{ex}\text{exp}[-\Delta G_\ddagger^\text{ex}(\Delta G_{OR}^\text{ex})/k_BT]$ is a partition function and $H_{OR}=\bra{O}\hat{H}\ket{R}$ is the electronic coupling.

The rate expression for electron transfer where $\ket{R}$ is a discrete state and $\{\ket{O}\}$ is a continuum of states can be derived using a generalized MHC integral, integrating over the continuum
\begin{equation}\label{genmhc}\begin{aligned}
k=\frac{2\pi}{\hbar} \, |H_{OR}|^2Q^{-1}\int d\epsilon_O \rho(\epsilon_O)n(\epsilon_O)\text{exp}\left(-\frac{\Delta G_\ddagger^\text{ex}(\Delta G_{OR}^\text{ex},\epsilon_O)}{k_BT}\right),
\end{aligned}\end{equation}
where $\rho$ is the density of the continuum, $n$ is the Fermi-Dirac occupancy, and $G^\text{ex}_\ddagger$ is now a function not only of some global shift $\Delta G_{OR}^\text{ex}$ but also the location of the initial state's energy $\epsilon_O$ within the continuum.
\begin{figure}[!hbt]
\includegraphics[width=1\columnwidth]{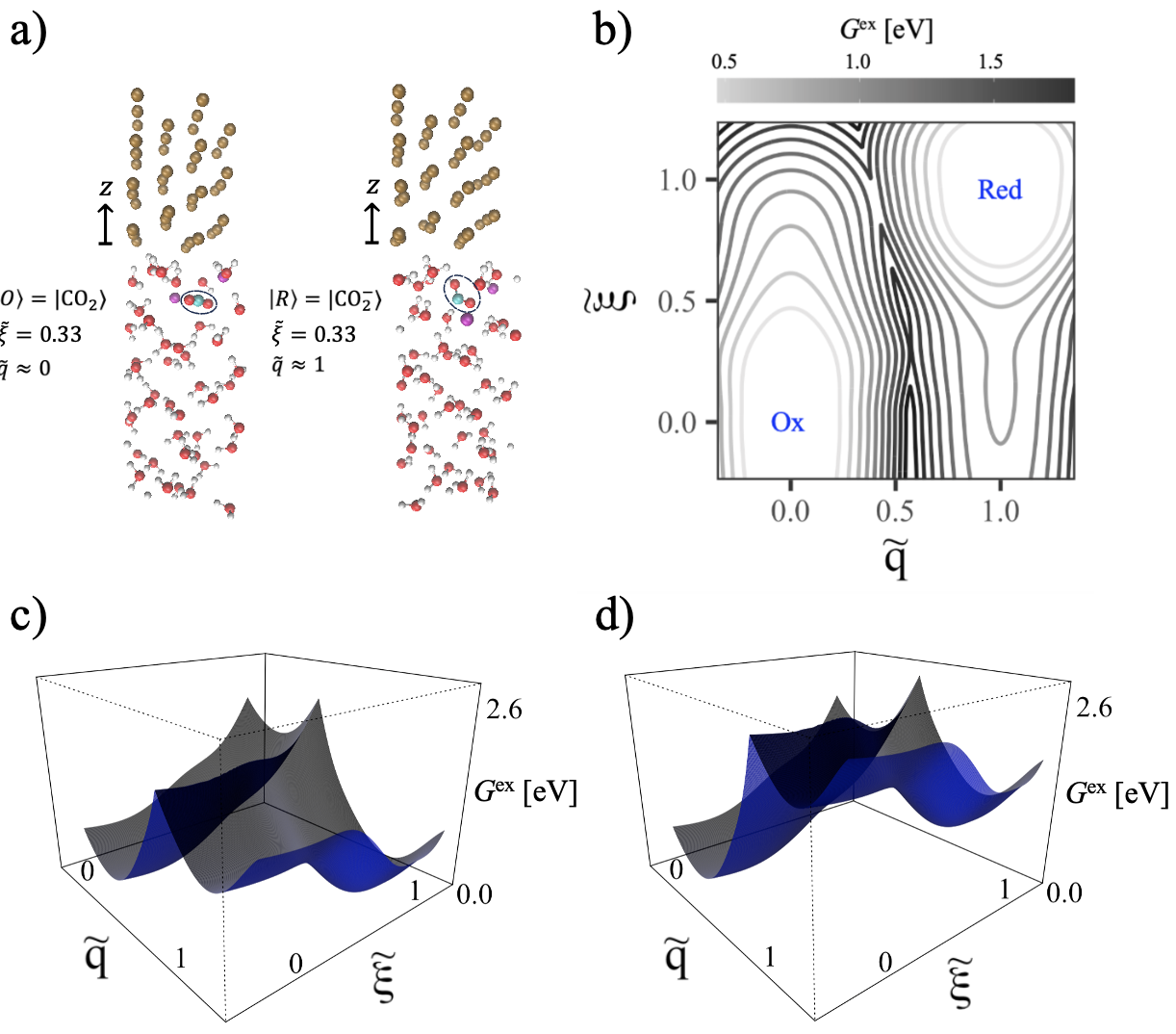}
\caption{Snapshots of the model aqueous CO$_2$–Au interface constrained to (left) the oxidized and (right) the reduced state. The circled CO$_2$ is linear in the oxidized state and bent in the reduced state. Au, C, O, H, and K are shown in gold, blue, red, white, and purple, respectively. The $z$-axis is taken normal to the electrode surface. For the snapshot shown, a harmonic restraint is used to fix $\tilde{\xi}\approx0.33.$ Similarly, the cDFT constraints fix $\tilde{q}\approx0,1$ in the left and right panels respectively. (b) Contour plot, and (c) perspective visualization of the 2D ground-state \textit{diabatic} free energy surface with thermodynamic driving force $\Delta G_{OR}^\text{ex}=0$. (d) Same as (c) except setting $\Delta G_{OR}^\text{ex}=$1 eV, which shifts the reduced diabat relative to the oxidized. The basins at points (0,0) and (1,1)---labeled "Ox" and "Red" in the contour plot respectively---are signatures of a CIET-like mechanism.}
\label{FIG1}
\end{figure}

{\it Adiabatic Corrections.} The formalism presented above is appropriate for the nonadiabatic (weak coupling) limit, where the electronic coupling is negligible compared to the reorganization energy \footnote{Formally, the degree of nonadiabatic vs. adiabatic character can be calculated using the Landau-Zener formula as described in Refs. \cite{Wittig2005LandauZener,Nitzan_2006}, adiabatic corrections to the landscape become significant. However, as we have argued in Ref. \cite{Abraham2025_lambda_eff}, adiabaticity does not alter the necessity of integrating over the band structure of the electrode. Therefore, calculation of the adiabatic surface as a function of $\Delta G_{OR}^\text{ex}$ is required.}. If an effective coupling $H_{OR}(\tilde{q},\tilde{\xi})$ were known, the adiabatic surface could be computed by diagonalizing the two-level system \begin{equation}\label{adiabatic_ciet}\begin{split}G^\text{ex}_{-}&(\tilde{q},\tilde{\xi},\Delta G_{OR}^\text{ex})=\frac{G_O^\text{ex}(\tilde{q},\tilde{\xi})+G_R^\text{ex}(\tilde{q},\tilde{\xi},\Delta G_{OR}^\text{ex})}{2}\\&-\frac{1}{2}\sqrt{(G_O^\text{ex}(\tilde{q},\tilde{\xi})-G_R^\text{ex}(\tilde{q},\tilde{\xi},\Delta G_{OR}^\text{ex}))^2+4|H_{OR}(\tilde{q},\tilde{\xi})|^2}.\end{split}\end{equation} Note that the true coupling value \begin{equation}H_{OR}(\textbf{R})=\bra{O(\textbf{R})}\hat{H}\ket{R(\textbf{R})}\end{equation} corresponding to a nuclear coordinate $\textbf{R}$ can be approximated from the Kohn-Sham cDFT orbitals according to Ref. \cite{Troy_mixed_cdft}, but will fluctuate throughout an \textit{ab initio} molecular dynamics (AIMD) trajectory. We can therefore time average to obtain a mean coupling
\begin{subequations}\label{effective_coupling}\begin{equation}
    H_{OR}(0,\tilde{\xi})=\frac{1}{T}\int dt\int d^{3N}\textbf{R}\delta(\tilde{\xi}(\textbf{R}_O(t))-\tilde{\xi})H_{OR}(\textbf{R}_O(t)),
\end{equation}\begin{equation}
    H_{OR}(1,\tilde{\xi})=\frac{1}{T}\int dt\int d^{3N}\textbf{R}\delta(\tilde{\xi}(\textbf{R}_R(t))-\tilde{\xi})H_{OR}(\textbf{R}_R(t)),
\end{equation}\begin{equation}H_{OR}(\tilde{q},\tilde{\xi})=H_{OR}(0,\tilde{\xi})+\tilde{q}[H_{OR}(1,\tilde{\xi})-H_{OR}(0,\tilde{\xi})],\end{equation}\end{subequations}  where $T$ is the total time of the trajectory, assumed to be sufficiently large. In Eq. (\ref{effective_coupling}) we have first averaged along the minima in the CV direction and then employed a linear interpolation to approximate the coupling for values $0<\tilde{q}<1$ (note that the linear approximation is only a starting point; more accurate dependencies of the coupling on $\tilde{q}$ can also be explored. In addition, alternative averaging schemes for the coupling, such as RMS averaging, may also be considered). As discussed in Ref. \cite{Abraham2025_lambda_eff}, the structure of the adiabatic rate expression differs only by the prefactor \begin{equation}\label{genmhc}\begin{aligned}
k^\text{adiabatic}=\nu\int d\epsilon_O \rho(\epsilon_O)n(\epsilon_O)\text{exp}\left(-\frac{\Delta G_\ddagger^\text{ex}(\Delta G_{OR}^\text{ex},\epsilon_O)}{k_BT}\right),
\end{aligned}\end{equation} where $\nu$ is the classical attempt frequency.

We emphasize that the formalism presented here has been kept general; the analytical CIET expressions of Refs. \cite{Bazant_CIET,Abraham2025_lambda_eff} can be derived upon specialization to a particular electrochemical system and making the assumption that the free energy surfaces have a particular shape.

{\it Example Application: CO$_2$ redox.} The choice of the CO$_2$ redox system for demonstration of the above methodology is motivated by the relative simplicity of approximating the diabatic states. Defining the diabatic states for lithium-ion intercalation, for example, is less straightforward because the lithium ion does not change its oxidation state during the intercalation process. Instead, the electrode is reduced to accommodate the lithium ion's charge, leading to diabatic states whose definition remains an open scientific question \cite{LIB_Science} and will be addressed in future work. Here, we instead demonstrate our methodology on the supposed rate-determining step of CO$_2$ reduction \cite{Qin2023_CO2_Reduction,zhang_driving_2020,Koper2024Theory}, where the diabatic states---$\ket{\text{CO}_2}$ and $\ket{\text{CO}_2^-}$---are obtained via simple constraints, and where a CIET-like approach is less explored.
\begin{table}[t]
\centering
{\footnotesize
\begin{tabular}{l|ccc|ccc}
\hline
\addlinespace[2pt]
$\Delta G_{OR}^\text{ex}$ &
$\Delta G_{\ddagger}^{\mathrm{CV,Red}}$ &
$\Delta G_{\ddagger}^{\mathrm{ET,Red}}$ &
$\Delta G_{\ddagger}^{\mathrm{2D,Red}}$ &
$\Delta G_{\ddagger}^{\mathrm{CV,Ox}}$ &
$\Delta G_{\ddagger}^{\mathrm{ET,Ox}}$ &
$\Delta G_{\ddagger}^{\mathrm{2D,Ox}}$ \\
\addlinespace[2pt]
\hline
\addlinespace[2pt]
$-0.5$ & 0.3 & 1.9 & 1.4 & 0.8 & 2.0 & 1.9 \\
 0.0 & 0.5 & 2.2 & 1.6 & 0.5 & 1.7 & 1.6 \\
 0.5 & 0.7 & 2.5 & 1.8 & 0.2 & 1.4 & 1.3 \\
 1.0 & 1.0 & 2.8 & 2.1 & 0.0 & 1.1 & 1.1 \\
\hline
\end{tabular}
}
\caption{\small Comparison of activation barriers (in eV) for CO$_2\rightarrow$CO$_2^-$ along the minimum energy path on the \textit{diabatic} surfaces using three approaches: (i) considering only the CV dimension (BV approach), (ii) considering only the ET coordinate (Marcus approach), and (iii) the full 2D landscape (CIET approach). Results are shown for different values of the thermodynamic driving force $\Delta G_{OR}^\text{ex}$.}
\label{tab:barriers_9x7}
\end{table}

Electrocatalyzed CO$_2$ reduction is of growing interest for various industrial processes and sustainability applications \cite{Science_CO2R,CO2R_review}, and gold and silver electrodes are noteworthy for their high product-specificity for CO \cite{CO2R_specificity}. Efficient ET in this system requires the presence of cations in the solution, and the mechanistic reason for this requirement has been debated \cite{cation_promoter,Ringe_cation,pH_cation}. Recent simulation results have suggested that the cations lower the barrier for CO$_2$ adsorption, thus facilitating the ET \cite{Qin2023_CO2_Reduction}. 

Consistent with this picture, we take as our classical CV $\xi$ the $z$-distance of the carbon atom from the electrode surface, and take $\ket{O}$ and $\ket{R}$ in Eq. (\ref{gap}) to be the electronic states $\ket{\text{CO}_2}$ and $\ket{\text{CO}_2^-}$ respectively. While previous formulations of CIET theory have suggested that the physically relevant CV is related to the distance from the electrode \cite{early_CIET,Bazant_CIET}, here we provide two microscopic justifications for this assumption. First, in Faradaic reactions the electronic coupling varies strongly with distance from the electrode, increasing substantially as the reacting molecule approaches the surface (see Fig. 2(a)) \cite{Abraham2025_lambda_eff}. As a result, the distance from the electrode captures the dominant variation in electronic coupling across the reaction landscape. Second, we find that this CV exhibits dynamics that are approximately orthogonal to the ET coordinate (i.e., weakly correlated fluctuations), and that alternative CV choices such as the O–C–O bond angle exhibit fluctuations strongly correlated with the ET coordinate (see Supplementary Material \cite{SM} for both analyses). Therefore, although other CVs may be considered, we find that this distance coordinate provides a physically meaningful parametrization of the CIET mechanism for this system.

Using this coordinate system, we compute the diabatic free-energy surface for CO$_2$ redox on a gold electrode in the presence of K$^+$ ions (see Fig. 1(a) and the Supplementary Material \cite{SM} for corresponding numerical details). Figure 1(b)--(d) show contour and perspective views of the resulting free energy surface for several thermodynamic driving forces $\Delta G^\text{ex}_{OR}$. From Eq. (\ref{bounds}) we obtain $(q_O,\xi_O)=(-7.1~\text{eV},5.1 ~\text{\AA})$ and $(q_R,\xi_R)=(5.5~\text{eV},2.4~\text{\AA})$, which map to the points $(0,0)$ and $(1,1)$ respectively upon nondimensionalization. The contour plot shows relative minima in both corners corresponding to the oxidized and reduced state---a signature of a CIET mechanism.
\begin{figure}
\includegraphics[width=1\columnwidth]{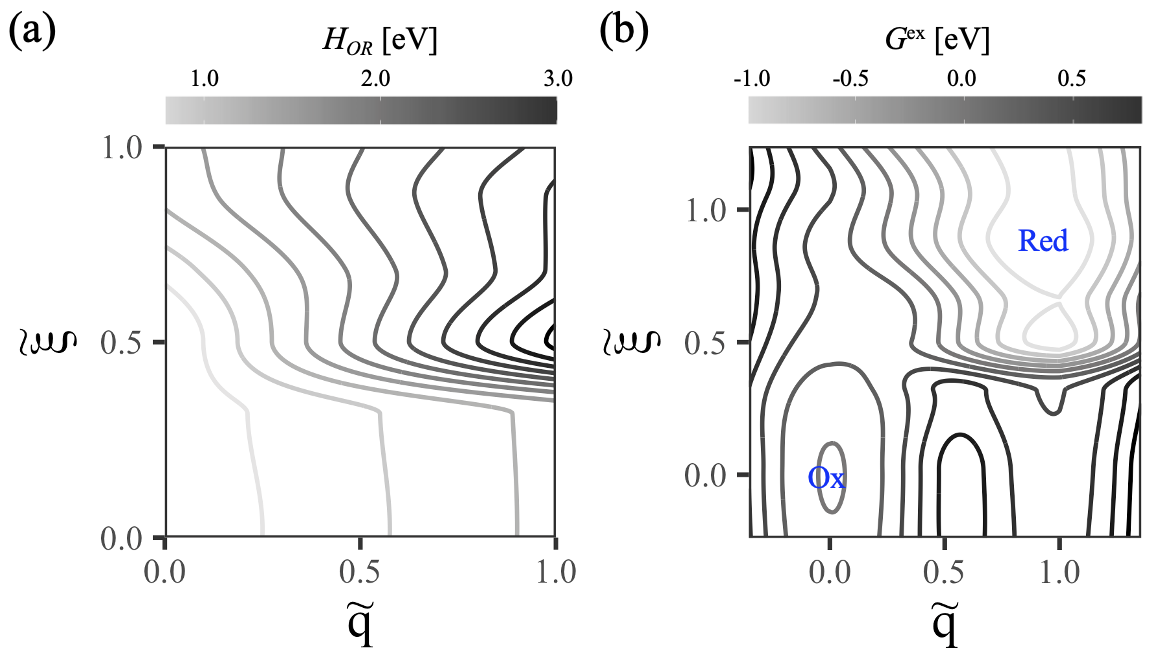}
\caption{(a) Contour plot of the electronic coupling for CO$_2$ redox across the 2D landscape. (b) Contour plot of 2D ground-state \textit{adiabatic} free energy surface setting $\Delta G^\text{ex}_{OR}=0$ eV.}
\label{FIG2}
\end{figure}

It is instructive to compare the barrier height in the diabatic 2D landscape with that obtained by considering only the CV or ET coordinate individually. This comparison can be done for both the reduction and oxidation directions, and for various thermodynamic driving forces $\Delta G_{OR}^\text{ex}$. Results of this comparison are shown in Table I. We denote by $G^{\text{CV,Red/Ox}}_\ddagger$ the reaction barrier that would be obtained by considering only the CV dimension, collapsing the ET dimension to the vertex of each Marcus parabola in the manifold. We denote by $G^{\text{ET,Red/Ox}}_\ddagger$ the reaction barrier that would be obtained by considering only the ET dimension, taking cross-sectional Marcus parabolas at fixed CV. Finally, we denote by $G^{\text{2D,Red/Ox}}_\ddagger$ the activation barrier obtained from the full 2D approach, which is obtained numerically from Eq. (\ref{ts}). Crucially, while $G^{\text{ET,Red/Ox}}_\ddagger$ corresponds to a valid (although in general higher-energy) reaction path, $G^{\text{CV,Red/Ox}}_\ddagger$ does not correspond to a physical reaction path satisfying the Franck-Condon principle \cite{Marcus_1956,Libby_Condon} and is included only for illustrative purposes. We see that the CV-only approach consistently underestimates the barrier for nonadiabatic ET in this reaction, suggesting the importance of ET. At the same time, the ET-only approach overestimates the barrier compared to the TS in the two-dimensional CIET landscape.

While the results reported so far pertain to the diabatic landscape, CO$_2$ electroreduction is thought to be strongly adiabatic \cite{Abraham2025_lambda_eff,Qin2023_CO2_Reduction,Koper2024Theory}, which we now show. Figure 2(a) shows a contour plot of the electronic coupling across the full 2D landscape, while Fig. 2(b) shows a contour plot of the computed adiabatic free energy surface adjusted according to Eq. (\ref{adiabatic_ciet}). Note that the coupling values are very large---on the order of 1--3 eV---in remarkable agreement with the theoretical prediction of Ref. \cite{Abraham2025_lambda_eff} and indicative of an adiabatic mechanism.
\begin{table}[t]
\centering
{\footnotesize
\begin{tabular}{ll|ccc|ccc}
\hline
\addlinespace[2pt]
 $\Delta G_{OR}^\text{ex}$ & $\Delta G_\text{eff}^\text{ex}$ &
$\Delta G_{\ddagger}^{\mathrm{CV,Red}}$ &
$\Delta G_{\ddagger}^{\mathrm{ET,Red}}$ &
$\Delta G_{\ddagger}^{\mathrm{2D,Red}}$ &
$\Delta G_{\ddagger}^{\mathrm{CV,Ox}}$ &
$\Delta G_{\ddagger}^{\mathrm{ET,Ox}}$ &
$\Delta G_{\ddagger}^{\mathrm{2D,Ox}}$ \\
\addlinespace[2pt]
\hline
\addlinespace[2pt]
$-0.5$ & $-1.8$ & 0.1 & 1.0 & 0.4 & 1.9 & 2.2 & 2.1 \\
 0.0 & $-1.4$ & 0.2 & 1.3 & 0.4 & 1.6 & 1.8 & 1.8 \\
 0.5 & $-1.0$ & 0.3 & 1.5 & 0.4 & 1.2 & 1.5 & 1.4 \\
 1.0 & $-0.6$ & 0.3 & 1.8 & 0.5 & 0.9 & 1.1 & 1.1 \\
\hline
\end{tabular}
}
\caption{\small Same as Table I, except using the ground-state \textit{adiabatic} surface. The quantity $\Delta G_\text{eff}^\text{ex}$ is defined as the free energy difference between the reduced and oxidized local minima of the adiabatic surface.}
\label{tab:barriers_9x7}
\end{table}

To calculate the TS within the adiabatic picture, we note that the TS is not confined to occur within the set $\mathcal{I}$ defined in Eq. (\ref{inter}), and so the restriction in the minimization in Eq. (\ref{ts}a) must be lifted. We therefore located the TS using a minimax (lowest-saddle) search, in which all neighboring grid points were connected by edges weighted by the higher of their energies, and the smallest edge-maximum path connecting the bottom-left and top-right minima was located via a Kruskal-style union–find algorithm \cite{kruskal1956}. Furthermore, note that in contrast to the nonadiabatic mechanism discussed above, the experimentally observed driving force $\Delta G_\text{eff}^\text{ex}$ is no longer strictly equivalent to the diabatic shift $\Delta G_{OR}^\text{ex}$ \cite{Abraham2025_lambda_eff}. 

The resulting adiabatic barriers are reported in Table II. We observe that many of the barriers are significantly lower compared to the diabatic landscape. Similar to the nonadiabatic case, we do not find consistent agreement between the CIET barrier and either one-dimensional approach.

We note that a standard approach for the adiabatic barrier is to calculate a ground-state free energy path along the CV coordinate \cite{Qin2023_CO2_Reduction,Axel_water}. While this standard approach may yield a qualitative picture of the reaction barrier, it provides only a single barrier corresponding to a fixed thermodynamic driving force. In contrast, the CIET framework enables explicit calculation of the full dependence of the activation barrier on the driving force, $\Delta G^\text{ex}_{\ddagger}(\Delta G^\text{ex}_{OR})$, which is required to evaluate electrochemical rate expressions and current–overpotential relationships \cite{Bazant_CIET,Abraham2025_lambda_eff}. This distinction is essential: the kinetics of electrochemical reactions are governed not by a single barrier, but by how that barrier evolves with applied potential over a band of electrode states. Furthermore, it is known that DFT-based methods often struggle with fractional charge states \cite{Perdew_fractional,Cohen_fractional}, whereas the integer charge states used in cDFT have been shown to help minimize this issue \cite{Troy_cDFT_proof}. For these reasons---in addition to the quantitative differences between the CIET and single-dimensional barriers reported---we expect that the methods presented in this Letter will provide a more accurate and physically applicable framework for computing electrochemical activation barriers. Future work will apply this formalism to predict experimental current–overpotential relations by approximating CIET parameters \cite{Bazant_CIET,Abraham2025_lambda_eff} from the \textit{ab initio} free energy landscapes.

{\it Acknowledgements.} This work was supported by the MIT Faculty Research Innovation Fund (FRIF) through the generosity of Irv and Melinda Simon. E.A. is grateful to Constantine Kyprianou, Ezra Alexander, Mohammadhasan Dinpajooh, Beck Hanscam, Ziyue Jiang, Matthias Kick, Abraham Nitzan, Joakim Stenlid, Leah Weisburn, Adam Willard, Junghyun Yoon, and Chao Zhang for useful discussions.

\bibliographystyle{aipnum4-1}

\end{document}